\documentclass[12pt, twoside]{article}
\usepackage[utf8]{inputenc}
\usepackage[english]{babel}
\usepackage{amsthm, amssymb, amsmath}
\usepackage[utf8]{inputenc}
\usepackage{graphicx}
\usepackage{geometry}
\usepackage{indentfirst}
\usepackage{booktabs}
\usepackage{setspace}
\usepackage{array}
\usepackage[boxed]{algorithm2e}
\usepackage[table]{xcolor}
\usepackage{longtable}
\usepackage{caption}
\usepackage{hyperref}
\usepackage{afterpage}
\usepackage{mathtools}
\usepackage{tikz}
\usepackage{nicefrac}
\usepackage{enumerate}
\usepackage{centernot}
\usetikzlibrary{arrows}
\graphicspath{ {./images/} }
\usepackage{relsize}
\usepackage[mathscr]{euscript}
\usepackage{float}
\usepackage{dblfloatfix}
\usepackage{url} 
\usepackage{cleveref}
\usepackage{subcaption}
\usepackage{cite} 
\usepackage{lineno}  

\newcommand{\rev}[1]{{\color{black}{#1}}}

\usepackage{color}
\usepackage[]{xcolor}

\usepackage{titlesec}
\titleformat{\section}[block]{\bfseries\MakeUppercase}{\thesection}{1em}{\fontsize{14}{16}\selectfont}
\titleformat{\subsection}[block]{\bfseries\fontsize{12}{14}\selectfont}{\thesubsection}{1em}{}
\titleformat{\subsubsection}[block]{\itshape\fontsize{12}{14}\selectfont}{\thesubsubsection}{1em}{}

\definecolor{lightgrey}{rgb}{0.83, 0.83, 0.83}
\begin{document}
\begin{centering}
{\Large{Identifying Spatiotemporal Patterns in Opioid Vulnerability: Investigating the Links Between Disability, Prescription Opioids and Opioid-Related Mortality}}\\[.5cm]

Andrew Deas\textsuperscript{1,3,*}, Adam Spannaus\textsuperscript{3}, Hashan Fernando\textsuperscript{2}, Heidi A. Hanson\textsuperscript{3}, Anuj J. Kapadia\textsuperscript{3}, Jodie Trafton\textsuperscript{4}, and Vasileios Maroulas\textsuperscript{1}\\[0.5cm]
\end{centering}

\begin{flushleft}
\textsuperscript{1}Department of Mathematics, University of Tennessee, Circle Dr, Knoxville, 37916, TN, USA\\

\textsuperscript{2}The Bredesen Center for Interdisciplinary Research and Graduate Education, University of Tennessee, Middle Dr, Knoxville, 37996, TN, USA\\

\textsuperscript{3}Computational Sciences and Engineering Division, Oak Ridge National Laboratory, Bethel Valley Road, Oak Ridge, 37830, TN, USA\\

\textsuperscript{4}Office of Mental Health and Suicide Prevention, Veterans Health Administration,Willow Road, Palo Alto, 94025, CA, USA\\[0.25cm]

*Corresponding author email: deasaj@ornl.gov\\
Contributing authors: spannausat@ornl.gov; hfernand@vols.utk.edu; hansonha@ornl.gov; kapadiaaj@ornl.gov; Jodie.Trafton@va.gov; vmaroula@utk.edu\\[0.5cm]
\begin{centering}
\textbf{Abstract}\\
\end{centering}
\textbf{Background:} The opioid crisis remains one of the most daunting and complex public health problems in the United States. This study investigates the national epidemic by analyzing vulnerability profiles of three key factors: opioid-related mortality rates, opioid prescription dispensing rates, and disability rank ordered rates.

\textbf{Methods:} This study utilizes county level data, spanning the years 2014 through 2020, on the rates of opioid-related mortality, opioid prescription dispensing, and disability. To successfully estimate and predict trends in these opioid-related factors, we augment the Kalman Filter with a novel spatial component. To define opioid vulnerability profiles, we create heat maps of our filter's predicted rates across the nation’s counties and identify the hotspots. In this context, hotspots are defined on a year-by-year basis as counties with rates in the top 5\% nationally. 

\textbf{Results:} Our spatial Kalman filter demonstrates strong predictive performance. From 2014 to 2018, these predictions highlight consistent spatiotemporal patterns across all three factors, with Appalachia distinguished as the nation’s most vulnerable region. Starting in 2019 however, the dispensing rate profiles undergo a dramatic and chaotic shift.\\

\textbf{Conclusions:} The initial primary drivers of opioid abuse in the Appalachian region were likely prescription opioids; however, it now appears that abuse is sustained by illegal drugs. Additionally, we find that the disabled subpopulation may be more at risk of opioid-related mortality than the general population. Public health initiatives must extend beyond controlling prescription practices to address the transition to and impact of illicit drug use.\\[1cm]


\end{flushleft}

\section{Introduction}

Propelled by prescription practices and the subsequent misuse of opioids, the United States continues to struggle with the opioid crisis \cite{boiling_point,three_waves_article,cdc_three_waves}. The crisis was ignited in the 1990s with the aggressive prescribing of potent opioids like OxyContin. Despite the potential for dependency and abuse, these drugs were marketed with the promise of being non-addictive which led to an increase in prescription opioid consumption and significantly contributed to the epidemic's progression \cite{boiling_point,three_waves_article,cdc_three_waves}. As the crisis continues to unfold, the medical community and regulatory bodies face the difficult task of balancing effective chronic pain management against the risks of opioid abuse \cite{lancet_article}. 

In efforts to find this balance, certain measures, like prescription drug monitoring programs \cite{pdmp}, have effectively reduced the national rate of opioid prescriptions by 44.4 percent from 257.9 million in 2011 to
143.4 million in 2020 \cite{cdc_od_epidemic,ama_report}. Despite this decline however, opioid-related overdose deaths have risen, likely driven by the increasing prevalence of illicit opioids such as fentanyl and heroin \cite{cdc_three_waves,fda}. Nevertheless, such statistics may belie the more insidious ongoing impact of prescription opioids. For example, data has suggested some patients can transition from medically prescribed opioids to illicit use or develop a substance use disorder \cite{nida_prescription_facts, fda}. Such findings highlight the complexity of the opioid epidemic and underscore the necessity of a continued investigation into the evolving role of prescription opioids \cite{vital_signs,poly_substance,opioid_regulation}.

The complexity of the opioid crisis is further compounded by the fact that certain subpopulations are more prone to opioid abuse than the general population, making it challenging to develop uniform prescription practices. For instance, research has indicated that individuals with disabilities are more likely to misuse opioids and develop opioid use disorders \cite{disabled_oud}. Yet, they are often less likely to receive adequate treatment compared to those without disabilities \cite{disabled_oud}. For the aforementioned reasons, our study chose to survey the national opioid crisis by exploring three key factors: opioid-related mortality rates, opioid prescription dispensing rates, and disability rank ordered rates. We sought to uncover insights into each individual factor, as well as their potential interrelationships.

For data estimation and prediction, the cornerstone of our analysis utilizes a Kalman filter \cite{og_kalman,understanding_kalman}. The Kalman filter uses a series of observations to estimate unknown parameters and has been applied in many diverse settings, from navigating astronauts to the moon to real-time vehicle tracking \cite{kalman_moon,kalman_applications}. \rev{Despite its versatility, such applications primarily focus on temporal statistical modeling,} as the method traditionally lacks a spatial framework. To overcome this limitation, we propose a novel approach to augmenting the Kalman filter with a spatial component. This yields a principled framework to capture the complex geographical interrelations among counties within the United State's diverse landscape, thereby improving the accuracy of our opioid-related data estimations and predictions.

To define opioid vulnerability profiles, we create heat maps of our filter's predicted rates across the nation’s counties and identify the hotspots. In this context, hotspots are defined on a year-by-year basis as counties with rates in the top 5\% nationally. Multiple contiguous hotspot counties are referred to as clusters of hotspots and indicate the most vulnerable areas in the nation. \rev{Although many studies have examined spatiotemporal variations in opioid-related outcomes, most have concentrated on specific states or localized areas, limiting their ability to capture national patterns and trends \cite{hotspots_virginia, hotspots_new_york, hotspots_rhode_island, hernandez_epidemic_response, hotspots_massachusetts, hotspots_ohio}. Additionally, much of this research draws from clinical data sources, such as hospital records or autopsy reports, which, while valuable, may not fully capture the broader public health dynamics at play. To address these gaps, we examine the opioid crisis from a nationwide public health perspective, leveraging county-level data to thoroughly investigate community-level vulnerabilities across the United States.}

This study utilizes rates of opioid-related mortality, opioid prescription dispensing, and disability aggregated at the county level spanning the years 2014 through 2020. The final year in the study is the latest point available for comprehensive data collection across all datasets, but also stands as a pivotal moment in public health due to the COVID-19 pandemic. Although the impact of COVID-19 on the opioid crisis and its disruption of data collection are outside the scope of this study, it is important to acknowledge that the challenges associated with the opioid crisis were notably exacerbated by the 2020 pandemic \cite{covid19_impact,counterfactual_covid,county_penn_covid}. Consequently, 2020 is a critical year for examination within the context of the opioid crisis. 

This paper begins by framing our study within the scope of existing research, followed by a detailed exposition of the data and methodologies employed. We then articulate our findings before delving into a discussion that situates these findings within the wider discourse on managing the opioid crisis and the potential implications for future public health strategies. The code to reproduce our results is available in the Github repository linked in the `Availability of data and materials' section.

\subsection{Related works}

The intertwined dynamics of opioid mortality, prescription rates, and social vulnerability has been explored in previous research \cite{correlation,prescr_patterns_death,pain_management,social_prescr_factors}. Such studies have concluded that the crisis's roots extend beyond prescription practices to broader socioenvironmental issues and advocate for addressing structural addiction determinants \cite{best_practices,no_easy_fix}. Specifically, factors such as disability, demographic status, lack of health insurance, and low income have been identified as significant predictors of opioid misuse and mortality \cite{svi_ml,sdoh,biopsychosocial}. Building upon such previous insights, our study offers novel geospatial evidence to the discussion.

The Kalman filter uses a series of observations to estimate unknown parameters and has been applied in many diverse settings from navigating astronauts to the moon to real-time vehicle tracking \cite{og_kalman,understanding_kalman,kalman_moon,kalman_applications}. In such Kalman filter applications, numerous methods have been innovated to integrate spatial components into the filter \cite{kriged_kf,snow_water,dimension_reduced,spatial_Kalman}. For example, one study integrated the Kalman filter with the Kriging method; while another utilized the Kalman filter in conjunction with a separate spatiotemporal model. Extending this line of innovation, our study incorporates a spatial component through the process covariance matrix. Within this matrix, an exponential decay function is used to model spatial correlations. Our novel and independently derived approach bears similarity to the one adopted by Rougier et al. \cite{similar_kalman} in 2022; they however utilized an alternative function to compute covariance.

Heat maps are widely recognized and utilized for their efficacy in visualizing data intensity across regions. They translate complex datasets into choropleth maps. These maps facilitate an intuitive grasp of the varied vulnerability levels across a geographical landscape \cite{heatmaps}. To further delineate such landscapes, hotspot identification is frequently used. Its utilization can play a pivotal role in epidemiology and public health research for pinpointing areas that are disproportionately impacted by various health issues \cite{what_is_a_hotspot,prenatal_women,worldwide_hotspots,mass_hotspots}. For example, advanced Bayesian models have identified Ohio's most vulnerable hotspots to opioid overdose mortality \cite{hotspots_ohio}, and spacetime random forest models have unraveled the complex geospatial patterns of opioid-related crime in Chicago \cite{drug_crime}. Our study highlights the public health utility of integrating the Kalman filter with both heat maps and hotspot identification.

\section{Methods}

\subsection{Data}

This study examines the rates of opioid-related mortality, opioid prescription dispensing, and disability across the United States from 2014 to 2020. It is important to note that each dataset was collected differently: the opioid-related mortality rates are measured per 100,000 persons, the opioid prescription dispensing rates are measured per 100 persons, and the disability rank ordered rates are measured as percentile ranks from 0 to 100. Throughout our study period, the county structure of the United States changed due to the formation of new counties and the reconfiguration of existing ones \cite{county_changes}. However, the Kalman filter requires a fixed state space for its predictions, so our data has been curated to reflect the 2020 county structure, which includes 3,143 counties. This step ensures that we provide the most up-to-date and cohesive representation of the national county landscape with respect to the data. \rev{Additionally, at most four counties were added or removed in any given year throughout the study period, meaning this curation to the 2020 structure had minimal impact on our overall results.}

The drug mortality data were sourced from HepVu \cite{mort_rates}, originally collected by the Centers for Disease Control and Prevention's (CDC) National Center for Health Statistics and the National Vital Statistics System. These data represent narcotic overdose deaths per 100,000 persons, classified according to the International Classification of Diseases, Tenth Revision (ICD-10) codes \cite{mort_methods}. It is important to understand that these rates serve as indicators of opioid misuse rather than exact counts of opioid overdose mortality. Also sourced from the CDC are the opioid dispensing rates; these data reflect the rates of retail opioid prescriptions dispensed per 100 persons per year \cite{disp_rates}. 

The disability rates utilized in this study are measured as percentile rank estimates of the civilian non-institutionalized population with a disability in each county. These data were sourced from the CDC and Agency for Toxic Substances and Disease Registry's Social Vulnerability Index (SVI) which utilizes data from the American Community Survey to assess the resilience of communities to external stresses on human health \cite{svi_rates}. Given the SVI's biennial publication, we impute data for the intervening years by calculating half the difference between consecutive biennial data points and adding it to the earlier year's rates. This method, with the understanding that yearly rate fluctuations are modest, provided a viable way to analyze the impact of disability rank ordered rates on the national opioid crisis throughout the entire study period. Notably, Rio Arriba County, New Mexico, experienced a data collection error in 2018 \cite{data_error}. Therefore, for this county, we employed a more granular approach by taking a quarter of the difference between the 2016 and 2020 data points and adding it consecutively to each year starting in 2016. 

For counties with missing annual data rates, we assigned a value of 0 to those counties. This approach was chosen over data imputation for several reasons. Firstly, the disability and mortality rate data had minimal missing values. The disability rate data had no missing values for any year in the study. While the mortality rate data was missing values for only eleven counties each year, representing just 0.003\% of the data annually. Moreover, the set of missing counties in the mortality rate data was consistent across all years.

The dispensing rate data, however, presented more challenges with missing values: 184 counties in 2014, 181 in 2015, 182 in 2016, 189 in 2017, 263 in 2018, 49 in 2019, and 62 in 2020. The early years of the study had a higher incidence of missing data, up to 8.3\%, but the situation improved significantly in 2019 and 2020, with missing data reduced to a maximum of 1.9\%. Data quality issues in opioid regulation have been previously documented as a significant problem \cite{prescription_data_quality}. Additionally, since the writing of this manuscript, the CDC webpage where the opioid dispensing data were originally obtained has been updated with data from a new source and no longer includes any data prior to 2019. This poses challenges to potential data handling methods; to ensure consistency, we decided to assign missing values in the prescription rate dataset to 0.

While assigning a value of 0 to missing data does introduce a bias, this bias is minimal in the context of our analysis. Our primary focus is on identifying the most vulnerable regions with the highest rates nationwide. By setting missing rates to 0, we exclude these counties from being identified as vulnerable. This approach does carry the risk of overlooking some potentially vulnerable counties due to missing data, but it ensures that we do not artificially inflate the vulnerability of a region. More importantly, it prevents genuinely critical areas with available data from being overshadowed.

\subsection{Kalman filter model}

The Kalman filter operates on two principal equations: the state update equation, which models the evolution of a system's state over time, and the observation equation, which links the system's true state to observed data. These enable the Kalman filter to refine its estimates by updating previous predictions with new data. Opioid-related county-level data inherently possess both geospatial and temporal dimensions, making our data well-suited for the Kalman filter. Its inherent temporal nature allows it to leverage the evolution of opioid-related data over time, uncovering temporal patterns and trends. The spatial dimension is incorporated through the filter's covariance matrix, capturing the spatial correlations between counties in the United States. A detailed mathematical exposition of the Kalman filter algorithm is provided in \cref{apdx_kalman_algo}.

We now discuss our framework for understanding the dynamics driving the opioid crisis and how it aligns with the Kalman filter. In this framework, we analyze opioid trajectories in the United States at the county level, treating the state of the nation in any given year $t$ as a random vector, $N_t\in\mathbb{R}^d$, where $d$ is the number of counties in the nation. The nation is modeled as evolving from its immediate past state, $N_{t-1}$, by the following equation:
\begin{equation}\label{eq:state_evolution}
N_t = N_{t-1}+\varepsilon_t,\quad \varepsilon_t\sim\mathcal{N}(0,Q).
\end{equation}
The normal random vector $\varepsilon_t$ represents the change in the nation's state from one year to the next. 

The process covariance matrix $Q$ is specifically designed to capture the spatial correlations between counties. The entries of $Q=[q_{ij}]$ are inversely proportional to their respective county's geographical centers of population \cite{pop_centroids} $x_i$ and $x_j$, computed through an exponential decay function:
\begin{align*}
    q_{ij}=\exp\{-b\cdot d(x_i,x_j)\}.
\end{align*} 
Here $q_{ij}$ denotes the correlation between two counties, $d(x_i,x_j)$ represents the geographical distance between $x_i$ and $x_j$ measured using the haversine function, and $b>0$ is the decay rate that modulates the speed at which correlation decreases with distance. This decay rate is tailored to each factor in the data independently; and ensures that the correlation $q_{ij}$ diminishes to 50\% at a predetermined distance threshold. This threshold is derived from the approximate diameter of the most visually pronounced vulnerable region identified in the heat map visualizations of the data. In this way, we ensure a higher correlation among proximate counties and a significantly reduced correlation for more distant ones. Further details on the parameter values utilized in our study are available in the Python code which constructs the covariance matrices for each dataset. This code can be found in the linked Github repository.

\rev{We selected the exponential covariance function over alternatives, such as the Gaussian, because of its ability to better capture the complex spatial patterns observed in our data. Opioid-related outcomes often exhibit abrupt changes between neighboring counties that a Gaussian function, which enforces smoother spatial transitions, may fail to capture. The exponential decay function accommodates these sharp transitions while preserving long-range correlations, offering a better fit for the non-smooth spatial variations seen in the heat maps. Additionally, the Gaussian function's light tails make it less effective at capturing rare events, which are important when modeling abrupt regional disparities in opioid-related outcomes.}

\rev{In constructing the covariance matrix, we used distances between county centroids rather than adjacency-based connections to better capture continuous spatial relationships across U.S. counties. Adjacency-based models define spatial correlation strictly through shared borders, imposing a binary structure where counties are either connected or not, potentially oversimplifying complex spatial dependencies. This can be particularly limiting for opioid-related outcomes, which may be influenced by broader regional dynamics extending beyond immediate neighbors. By leveraging centroid distances between county population centers, our approach models gradual changes in spatial correlation across varying distances. Moreover, given the significant heterogeneity in county sizes and shapes across the United States, centroid-based distances provide a consistent and scalable measure of proximity.}

Our observed data, i.e., the rates of opioid-related mortality, prescriptions and disability in year $t$, can be modeled, with some degree of error, as measurements of the true underlying state. Thus, letting $D_t$ symbolize the observed data for an opioid-related variable in year $t$, the observation can be modeled as:
\begin{equation}\label{eq:measurement_eqn}
D_t = N_t + \eta_t,\quad \eta_t\sim\mathcal{N}(0,R).
\end{equation}
The normal random vector $\eta_t$ represents the measurement error in data collection, reporting, or other discrepancies from the true state. $R$ is the observation covariance matrix and quantifies the uncertainty in the data measurements. To choose a data uncertainty level for our study, we conducted a sensitivity analysis. This analysis examined various uncertainty levels, specifically 1\%, 3\%, and 5\%, by evaluating the respective changes in the model's predictive performance. Optimal predictive accuracy was found with a 1\% data uncertainty level. Consequently, a 1\% uncertainty assumption was adopted for all data measurements throughout our study. The results of this sensitivity analysis are summarized in \cref{apdx_uncertainty}.

We employed the Kalman filter to model the dynamics of the opioid crisis by selecting \cref{eq:state_evolution} as the state update equation and \cref{eq:measurement_eqn} as the observation equation. This model allows for continual updates to predictions as new data becomes available, which is vital for addressing the complex and evolving nature of the opioid epidemic. By enabling timely identification of emerging trends and patterns, this approach can inform and evaluate public health strategies and interventions.

\subsection{Heat maps and hotspot identification}

To construct heat maps of the distribution of rates across the nation's counties and identify the hotspots, we leverage the filter's predictions in each year by calculating the cumulative distribution function values for each county. These values are then used to categorize the counties into 20 distinct vulnerability levels, increasing in evenly spaced 5\% intervals. This results in a color gradient on the heat maps, transitioning from dark blue for the least vulnerable counties with the lowest rates to dark red for the most vulnerable counties with the highest rates. The most vulnerable counties, i.e., the counties whose predicted rates surpass the 95th percentile of the fitted normal distributions, are identified on a year-by-year basis and defined as the hotspots. Multiple contiguous hotspot counties are referred to as clusters of hotspots and indicate the most vulnerable areas in the nation.

\section{Results}

\subsection{Efficacy analysis of the filter's performance in the 2020 prediction year}

Presented here are the results for our spatial Kalman filter which was initialized with 2014 data and trained on data from 2015 to 2019 before generating predictions for 2020. Training, in this context, means allowing the filter to refine its estimates using the observed data; while its estimates were made without further refinement in the 2020 prediction year. We chose to predict for only a single year to prevent the accumulation of large errors inherent in multi-year forecasting. Long-term forecasting is accommodated by the filter's design however, and investigated in \cref{apdx_told_comparison}. In this investigation, the efficacy of multi-year predictions from the filter is examined by comparing its performance when trained on progressively less data to that of the fully trained filter discussed here. \rev{In \cref{apdx_initializations_years}, we explore how shifting the initialization year affects the model’s predictions, assessing whether more recent training data improves forecasting accuracy.} Presented in this section are the efficacy results for the filter's 2020 predictions. For the remaining years in the study, the efficacy metrics are summarized in \cref{apdx_training_efficacy}. In addition, the corresponding accuracy maps and error histograms can be found in the supplementary materials. 

The efficacy of our spatial Kalman filter in predicting county-level rates is quantitatively assessed using both accuracy and error metrics on a national scale. The accuracy metric results are visualized using accuracy maps; while error histograms are used to display the results of the error metrics. The metric used to calculate the error for each county is the absolute residual. The largest absolute residual, stemming from the filter's worst annual prediction, is then used to assess general accuracy in the following way: each county's error is normalized by the highest annual error observed nationwide. In addition to general accuracy, we also compute hotspot accuracy. Hotspots are defined on a year-by-year basis as counties with rates in the top 5\% nationally. The number of hotspots correctly predicted by the filter divided by the number of actual hotspots is defined as the hotspot accuracy.

\begin{figure*}[!h]
    \centering
    \includegraphics[width=\linewidth]{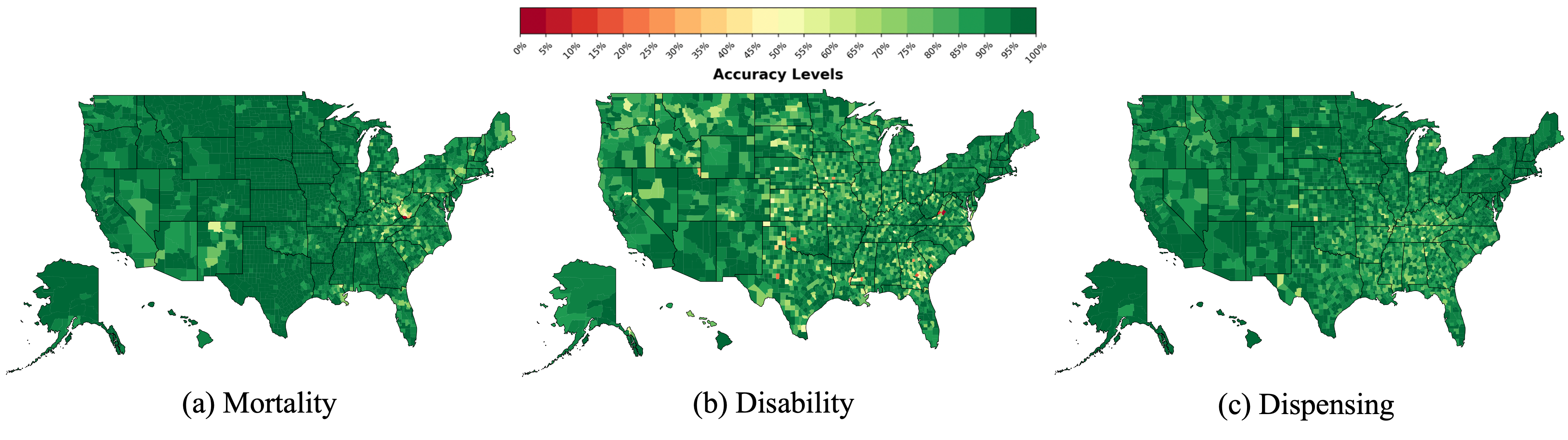}
    \caption{General accuracy maps for the 2020 (a) mortality, (b) disability, and (c) dispensing rate predictions. The maps are color-coded to represent increasing intervals of accuracy by 5\%, starting from dark red for the least accurate predictions, and progressing to dark green for the most accurate predictions. Figures are best viewed online in color.}
    \label{fig_accuracy_maps}
\end{figure*}

\begin{figure*}[!h]
    \centering
    \includegraphics[width=\linewidth]{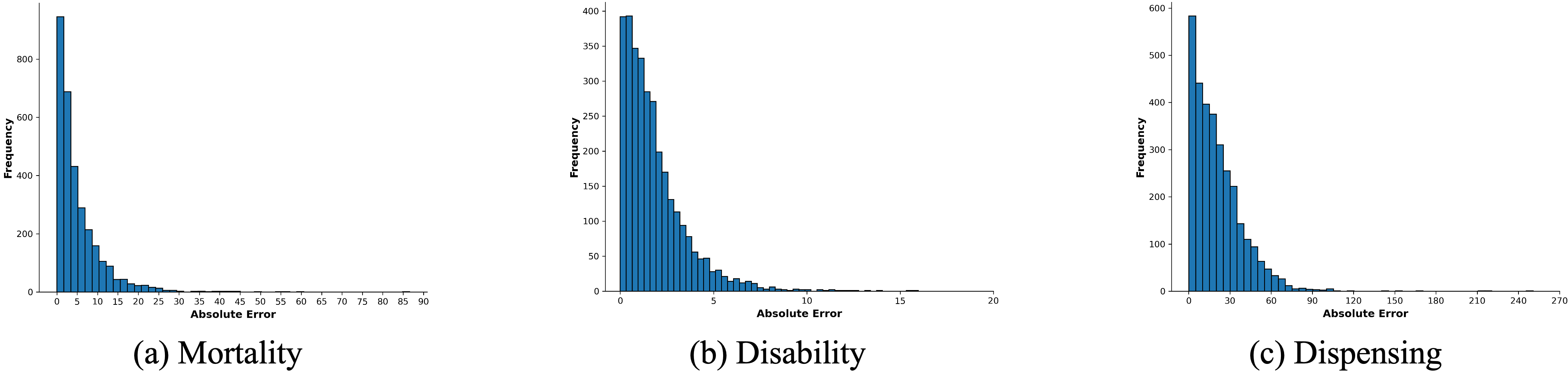}
    \caption{Histograms displaying the distribution of the filter's absolute errors for the 2020 (a) mortality, (b) disability, and (c) dispensing rate predictions.}
    \label{fig_err_histos}
\end{figure*}

\begin{figure*}[!h]
    \centering
    \includegraphics[width=\linewidth]{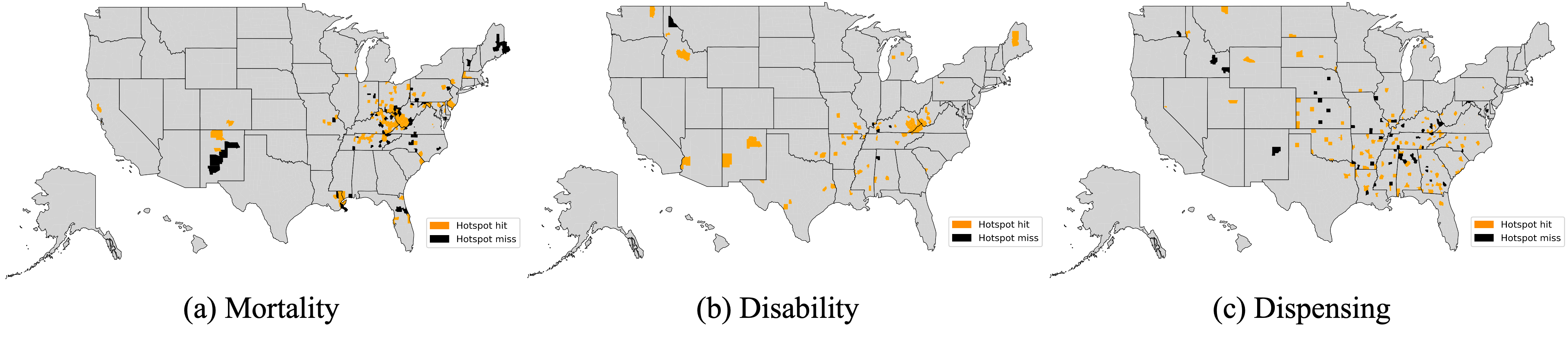}
    \caption{Hotspot accuracy maps for the 2020 (a) mortality, (b) disability, and (c) dispensing rate predictions. Orange colored counties represent accurately predicted hotspots, whereas those in black mark hotspots missed by our predictions. Figures are best viewed online in color.}
    \label{fig_hotspot_acc_maps}
\end{figure*}

For opioid-related mortality rates, measured per 100,000 persons, the filter achieved an average general accuracy of 94.00\%. The national accuracy distribution is displayed in \Cref{fig_accuracy_maps}. This map is characterized predominantly by green hues, indicating high predictive accuracy across the nation. A distinct region of lower accuracy, marked by yellow and red hues, is found in the southern part of West Virginia. This region experienced significant jumps in mortality rates from 2019 to 2020. These large jumps were difficult for the filter to predict and led to these distinguishable errors. The range of errors for this dataset, shown in \cref{fig_err_histos}, illustrates that our model's mortality rate estimates can deviate from the actual rates by an average of 5.20 deaths per 100,000 persons, with a maximum deviation of 86.61 deaths per 100,000 persons in the worst case. Additionally, for this dataset, the filter achieved a hotspot accuracy of 70.00\%. The hotspots which were correctly identified and those that were missed are shown in \cref{fig_hotspot_acc_maps}.


For the disability rank ordered rates, measured as percentile ranks from 0 to 100, our spatial Kalman filter demonstrated excellent predictive performance. The range of errors, shown in \cref{fig_err_histos}, demonstrates that our model's disability rank ordered rate estimates can deviate from the actual rank ordered rates by an average of 1.84 percentile rank points, with a maximum deviation of 15.97 percentile rank points in the worst case. This efficacy of the filter is also highlighted by its high hotspot accuracy of 94.19\%. The correctly predicted hotspots and those that were missed are shown in \cref{fig_hotspot_acc_maps}. On the other hand, the average general accuracy score of 88.49\% may not fully convey the filter's effectiveness on this dataset. This is because the maximum annual error of 15.97 used for normalization on this dataset results in even relatively small errors appearing more pronounced. This is visually evident in \cref{fig_accuracy_maps}; where we observe more yellow tones, denoting this magnification of smaller errors.

For the prescription dispensing rates, measured per 100 persons, the filter's predictions attained an average general accuracy of 91.61\%. The national accuracy distribution is visualized in \cref{fig_accuracy_maps}. This map is predominantly green; but upon close inspection, two red counties stand out: Union, South Dakota, and Montour, Pennsylvania. These counties experienced an extraordinary increase of over 350\% in their dispensing rates from 2018 to 2019. Based on this surge, the filter predicted similarly high increases from 2019 to 2020. However, the increases in 2020 were far less dramatic, leading to these significant prediction errors. The range of errors for this dataset, shown in \Cref{fig_err_histos}, illustrates that our model's dispensing rate estimates can deviate from the actual dispensing rates by an average of 21.02 prescriptions per 100 persons, with a maximum deviation of 250.67 prescriptions per 100 persons in the worst case. Additionally, the filter's hotspot accuracy on this dataset was 68.64\%. The correctly identified hotspots and those that were missed are shown in \cref{fig_hotspot_acc_maps}.

\rev{For context on the model’s performance across the different datasets, \cref{table_data_summary_statistics} in \cref{apdx_data_summary} presents the summary statistics for each factor across the entire study period. These statistics highlight the differences in data variability within each dataset, which can help explain the observed differences in model efficacy across the different datasets. The disability rates, being percentile rank ordered, form a more uniform and less variable dataset. In contrast, the prescription dispensing rates are inherently more chaotic. As shown in \cref{table_data_summary_statistics}, the dispensing rates consistently exhibit the highest values and standard deviations each year. The variability within the mortality rates dataset falls somewhere between these two extremes, it displays more structure than the chaotic dispensing rates but lacks the ordered nature of the rank-ordered disability rates. These inherent differences in data distribution and variability likely contribute to the observed variation in model performance across the three factors.}

\subsection{National opioid vulnerability profiles}

In this section, we present the vulnerability profiles uncovered by our spatial Kalman filter's predictions. \rev{Before proceeding to the results, it is important to clarify how the predictions used to generate these vulnerability profiles were produced. For the training years (2014–2019), the spatial Kalman filter updates its forecasts by incorporating the actual data from each year, resulting in calibrated estimates. In contrast, the 2020 predictions represent raw forecasts, as the model generates them solely based on patterns learned from the 2015–2019 training period, without access to 2020 data for adjustment. In all cases, yearly data were fed to the model sequentially.}

\begin{figure*}[!h]
    \centering
    \includegraphics[width=\linewidth]{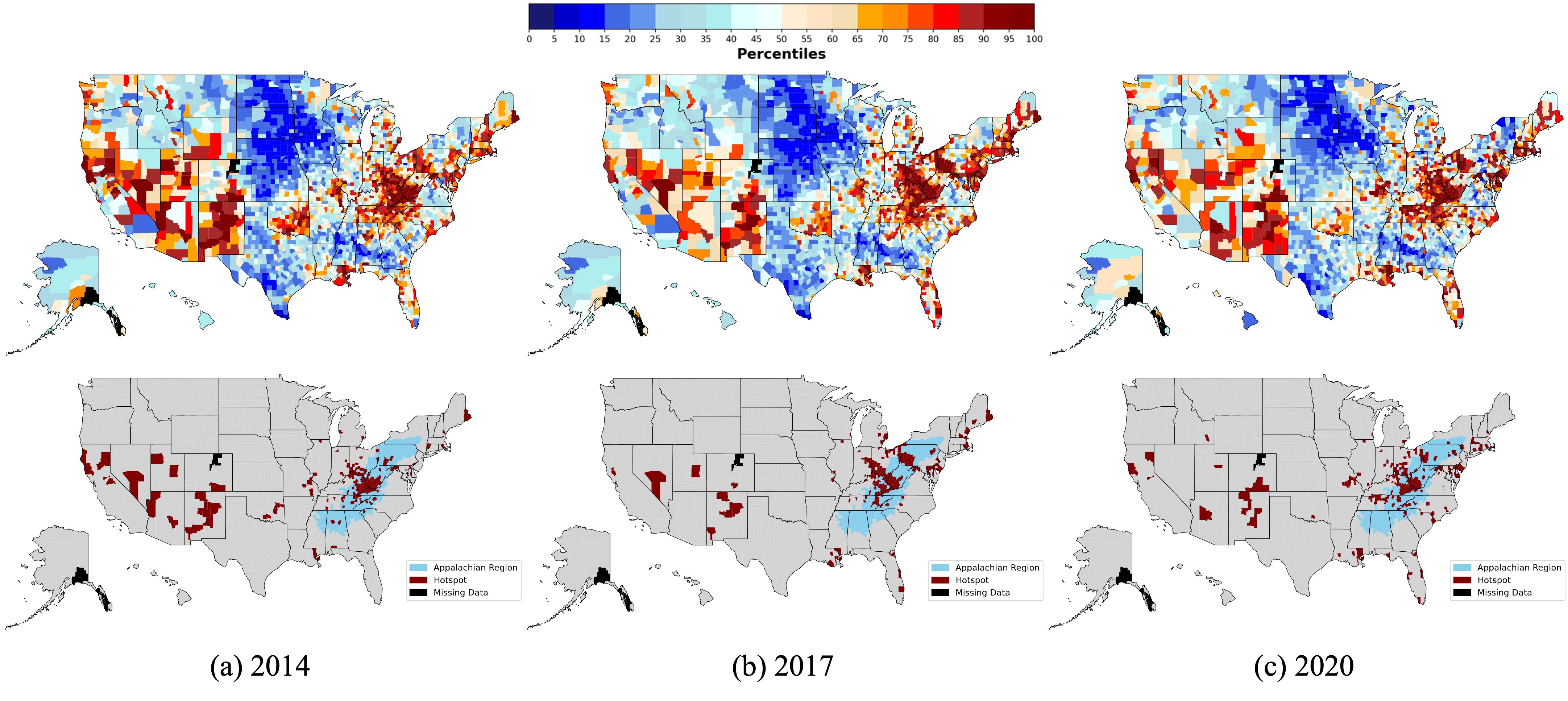}
    \caption{\rev{Heat maps and hotspot maps depicting the spatial Kalman filter’s predicted opioid-related mortality rates for (a) 2014, (b) 2017, and (c) 2020. These maps represent model-generated estimates rather than observed data. Counties with missing data are colored in black. Figures are best viewed online in color.}}
    \label{fig_od_maps}
\end{figure*}

\begin{figure*}[!h]
    \centering
    \includegraphics[width=\linewidth]{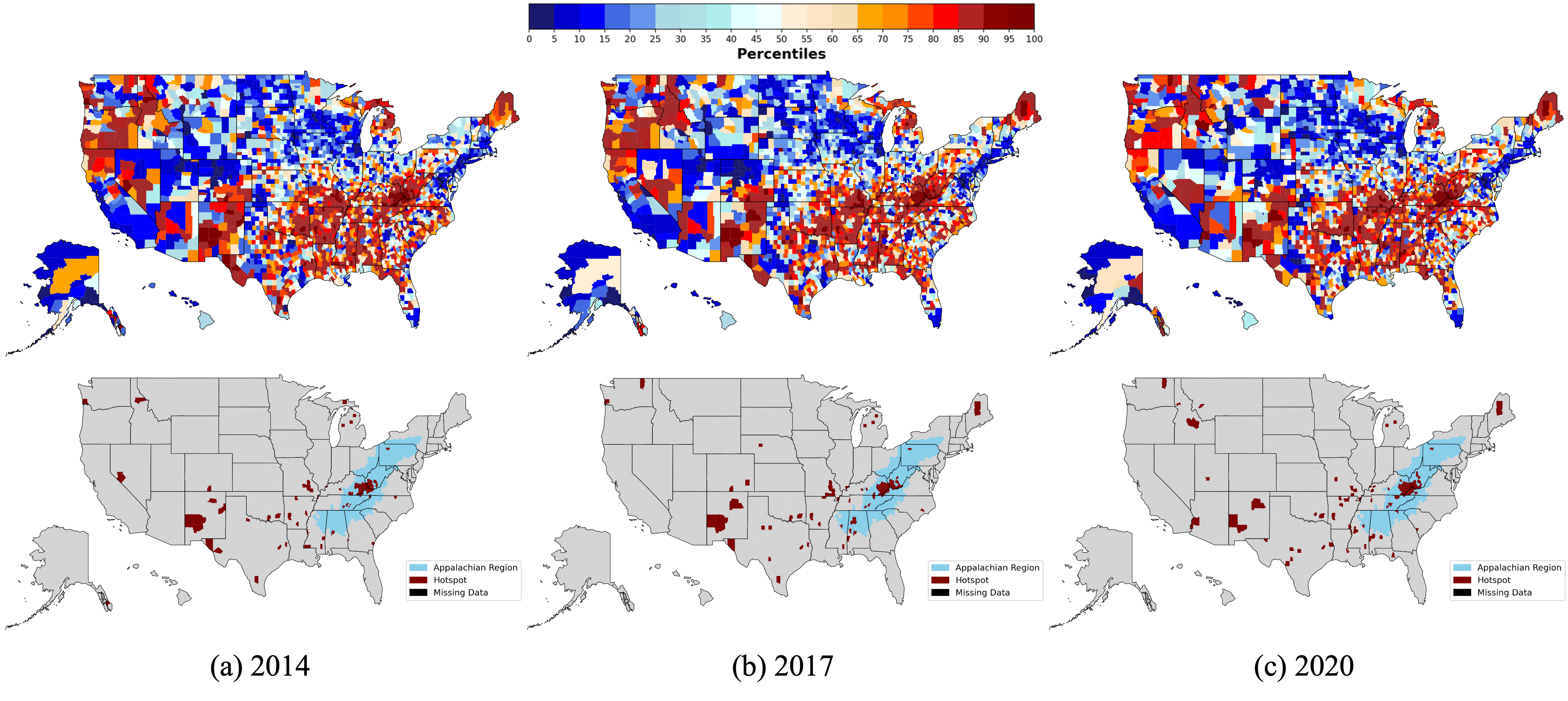}
    \caption{\rev{Heat maps and hotspot maps depicting the spatial Kalman filter’s predicted disability rank-ordered rates for (a) 2014, (b) 2017, and (c) 2020. These maps represent model-generated estimates rather than observed data. Counties with missing data are colored in black. Figures are best viewed online in color.}}
    \label{fig_dis_maps}
\end{figure*}

\begin{figure*}[!h]
    \centering
    \includegraphics[width=\linewidth]{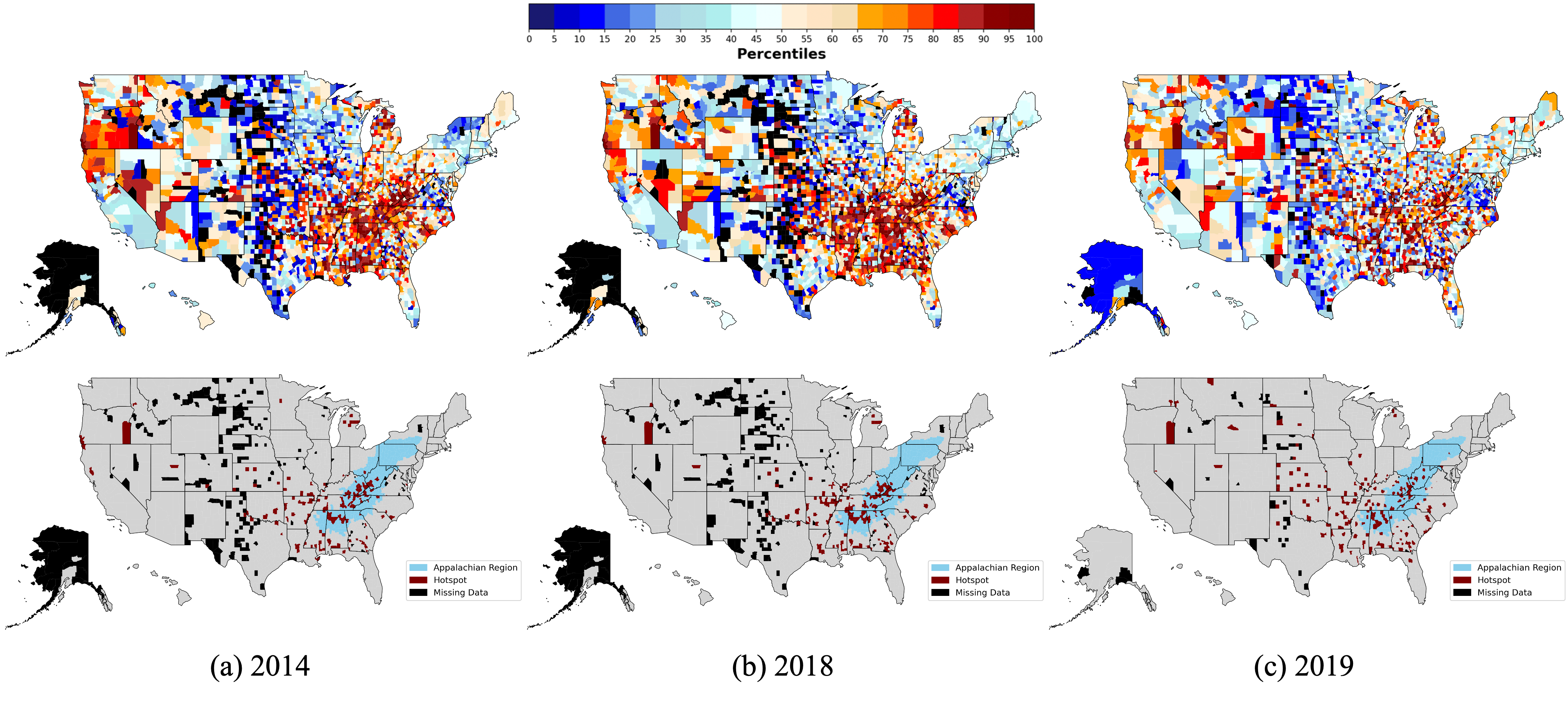}
    \caption{\rev{Heat maps and hotspot maps depicting the spatial Kalman filter’s predicted prescription opioid dispensing rates for (a) 2014, (b) 2018, and (c) 2019. These maps represent model-generated estimates rather than observed data. Counties with missing data are colored in black. Figures are best viewed online in color.}}
    \label{fig_dr_maps}
\end{figure*}

The heat maps and hotspot maps for the rates of mortality, disability and dispensing are showcased in \cref{fig_od_maps} through \cref{fig_dr_maps}, respectively. To illustrate the temporal evolution of each factor, the figures present the most significant maps. For both the mortality and disability rate predictions, we highlight maps from the years 2014, 2017, and 2020, while for the dispensing rate predictions, we focus on the years 2014, 2018, and 2019. Maps for the remaining years in the study can be found in the images folder of the supplementary materials.

The vulnerability profiles for the opioid-related mortality rates seen in \cref{fig_od_maps} exhibit consistent spatiotemporal patterns. Across these profiles, widespread vulnerability is consistently found with pronounced differences between regions. In both the Midwest and Northern Great Plains regions, there are recurring pockets of lower mortality rates, suggesting areas of resilience. Conversely, the Appalachian, Southwestern, Atlantic, and Gulf Coast regions exhibit pronounced vulnerability each year. Within these regions, Appalachia, highlighted in blue on each hotspot map, consistently emerges as a critical area of concern. Year after year, Appalachia contains the most visually evident cluster of hotspots.

The vulnerability profiles for the disability rank-ordered rates seen in \cref{fig_dis_maps} also exhibit consistent spatiotemporal patterns. Across these profiles, the Northern Great Plains, Pacific Southwest, and pockets of the Atlantic Coast repeatedly display notably lower rates of disability, indicating regions of relative resilience. In contrast, significant vulnerable areas are consistently found primarily in the Pacific Northwest, Southwest, Southeast, and Appalachian regions. Within these regions, the hotspots are dispersed each year but Appalachia repeatedly contains the most visually evident cluster. This strongly suggests that the Appalachian region not only contains the nation's most vulnerable counties to opioid-related mortality rates, but also to the disability rank-ordered rates. 

The vulnerability profiles for the prescription dispensing rates seen in \cref{fig_dr_maps} reveal a more complex pattern. In 2014, although we see some clusters of contiguous counties with high rates in the Pacific Northwest and Southeast, the counties manifesting the highest rates nationwide are primarily found in the Appalachian region. This is seen in the 2014 hotspot map and marks Appalachia as the nation's most vulnerable region in the first year of our study. The pattern seen in the 2014 profile persists with only slight changes until 2018, however, beginning in 2019 the pattern changes significantly. It is characterized by a chaotic nature with hotspots erratically spread across the nation and no visually noteworthy clusters of contiguous vulnerable or resilient counties. Additionally, starting in 2019, we see a noticeable drop in dispensing rates nationwide and the clusters of hotspots which distinguished the Appalachian region's vulnerability dissolve. 

\section{Discussion}

Leveraging the accurate predictions from our spatial Kalman filter, we analyzed the spatiotemporal patterns of the opioid crisis for the years 2014 through 2020 via constructed vulnerability profiles for three key factors: opioid-related mortality rates, opioid-dispensing rates, and disability rank ordered rates. \rev{Before analyzing these results in detail, it is important to validate our modeling approach by comparing the spatial Kalman filter to its generic counterpart. To ensure a fair evaluation, we assessed absolute errors rather than accuracy-based metrics. Accuracy measures can be disproportionately influenced by extreme values, meaning a model with a single large maximum error could misleadingly appear to have higher accuracy despite performing worse overall. By focusing on absolute errors, we provide a more stable and reliable comparison between the two models.}

\rev{Across nearly all datasets and years, our spatial Kalman filter outperformed the generic Kalman filter, yielding lower absolute errors in every case except for the 2020 predictions of prescription dispensing rates. This exception is notable because the dispensing rates dataset exhibited substantial variability and lacked a clear spatial structure, perhaps making it less suited for a spatially informed modeling approach. The spatial Kalman filter is designed to capture underlying geographic correlations, and its advantage is most pronounced when such patterns exist. This suggests that while the spatial Kalman filter is a powerful tool for modeling spatiotemporal processes, its effectiveness may depend on the presence of meaningful spatial patterns in the data.}

\rev{\Cref{fig_generic_comparison} in \cref{apdx_generic_comparison} presents a comparison of error distributions for four years in the study period, demonstrating the superior performance of our spatial Kalman filter in all but the case discussed above. Additional histogram comparisons for the full study period can be found in the accompanying GitHub repository. Overall, our spatial Kalman filter proves to be a more effective modeling approach than its generic counterpart. By integrating both temporal and geographic dependencies, the spatial Kalman filter enables more precise estimations, making it a valuable tool for studying complex datasets with inherent spatial patterns. The methodology we present can be broadly applied to other datasets where spatial relationships play a critical role and has broader utility beyond the opioid-related outcomes analyzed in this study.}

\rev{Having established the spatial Kalman filter as a more effective framework than its generic counterpart, we now turn to the opioid vulnerability profiles that emerged from its predictions.} The analysis of profiles for the mortality rates revealed widespread national vulnerability where the Appalachian region was repeatedly distinguished as the nation's most vulnerable area to opioid-related mortality rates. Correspondingly, the analysis examining the disability rank ordered rate profiles also uncovered widespread vulnerability where Appalachia was again repeatedly pinpointed as the nation's most vulnerable area to disability rank ordered rates. This alignment in the patterns for both disability and mortality profiles suggests that amidst the general fabric of the opioid crisis, the disabled subpopulation may be more at risk of opioid-related mortality than the general population. This finding corroborates previous work \cite{svi_ml,best_practices,appalachia}, and underscores the disproportionate outcomes of the opioid crisis. 

Additionally, the dual identification of Appalachia as the nation's most vulnerable area to both disability and mortality rates highlights the pressing need for support in this region. Appalachia faces significant challenges regarding resources available for disabled individuals. This region has a lower supply of healthcare professionals compared to the United States as a whole, including primary care physicians, mental health providers, specialty physicians, and dentists \cite{health_in_appalachia}. One possible public health intervention to mitigating the disproportinate impact on Appalachia could focus on improving access to healthcare and disability services. This could include increasing the number of healthcare providers, integrating disability support services into primary care, and ensuring that residents have access to rehabilitation and support programs. Additionally, policies aimed at addressing some the underlying social determinants of health, such as lower median household incomes and higher poverty rates, which are also prevalent in Appalachia, could significantly improve the health and well-being of disabled individuals in the region \cite{health_in_appalachia}.

Contrasting sharply with the patterns observed for the disability and mortality rates, our predicted prescription dispensing rates revealed more complex vulnerability profiles. From 2014 through 2018, these profiles were characterized by a consistent pattern which clearly distinguished Appalachia as the nation's most vulnerable region. While, beginning in 2019, the hotspot clusters marking the Appalachian region's vulnerability dissolved and we saw a noticeable drop in national dispensing rates. These changes are likely the result of several significant pieces of legislation addressing prescription opioids which were enacted in 2019. One notable example is the John S. McCain Opioid Addiction Prevention Act \cite{opioid_prevention_act}, which aimed to combat opioid addiction and reduce the over-prescription of opioids by limiting initial opioid prescriptions for acute pain to a seven-day supply. 

Starting in 2019, the dispensing profiles were not marked by any visually discernible spatial patterns, lacked notable clusters of hotspots, and generally exhibited more resilience than vulnerability across the nation. This suggests that current difficulties with opioid prescription practices are dispersed and not restricted to localized areas. Additionally, the observed resilient landscape implies that current prescription practices may not be the immediate focal point of concern. In fact, the lack of alignment between the vulnerability profiles of prescription and mortality rates indicates that prescription opioids might not be the primary current drivers of opioid-related fatalities nationwide. A conclusion that aligns with previous research \cite{no_easy_fix,best_practices}, and is additionally bolstered by contrasting national trends observed during our study period: despite a sharp 52.84\% drop in dispensing rates from 2014 to 2020, opioid mortality surged by 73.73\%. 

At the beginning of our study period, from 2014 to 2018, the vulnerability profiles for opioid-related mortality rates, opioid-dispensing rates, and disability rank-ordered rates all consistently identified the Appalachian region as the nation's most vulnerable area. However, beginning in 2019, the dispensing profiles showed a stark change with the previously observed vulnerability in Appalachia dissolving. This shift occurring at the end of our study period suggests two important insights. First, the initial primary drivers of opioid abuse in the Appalachian region were likely prescription opioids; then as more stringent legislation reduced the number of opioids dispensed, dependence likely shifted to illegal drugs to sustain the existing opioid abuse. Second, although a critical alignment between disability and mortality rate profiles was evident throughout the entire study period, the dramatic change in the dispensing profiles at the end of the study likely indicates that a connection between these two factors is no longer found in the realm of prescription opioids. Instead, this connection has also likely shifted to illicit drug use.

Recognizing that the patterns seen in the Appalachian region may reflect broader national trends and that the experiences of the disabled subpopulation could mirror those of other vulnerable groups, our study highlights the need for comprehensive and adaptable intervention strategies. Public health initiatives must extend beyond controlling prescription practices to address the possible transition to and impact of illicit drug use. Such initiatives could include developing integrated healthcare programs that combine support services with addiction treatment, and strengthening community-based programs aimed at substance abuse prevention by providing early support to at-risk individuals.

\section{Conclusion}

\rev{In this study, we leveraged the predictions from our spatial Kalman filter to investigate opioid vulnerability profiles and identify critical spatiotemporal patterns. Our findings highlight the potential of this approach for capturing geographic dependencies in opioid-related outcomes and documenting shifts in vulnerability over time. While our model was specifically applied to opioid-related mortality, prescription dispensing, and disability rates, the methodologies developed here can be adapted for a broader range of applications involving spatial and temporal dynamics.}

\rev{Our current framework models opioid-related mortality, prescription dispensing, and disability rates as independent processes, capturing temporal trends and spatial dependencies within each outcome separately. While incorporating socioeconomic covariates—such as poverty levels, healthcare access, or employment rates—could enhance interpretability, doing so within the current framework would model each outcome independently as a function of these factors, without explicitly capturing their interdependencies. A more comprehensive approach would involve a multivariate framework, in which socioeconomic factors serve as shared covariates across all three outcomes, enabling a more holistic analysis of their relationships. Future work could explore multivariate spatial Kalman filters, extended Kalman filters, or hierarchical modeling approaches to jointly model these interdependencies and better capture the complex interactions driving opioid vulnerability \cite{bayesian_heirarchal_approaches}.}

\rev{Finally, while our spatial Kalman filter effectively captures geographic proximity, temporal trends, and opioid-related outcomes, it does not explicitly adjust for external confounders such as policy interventions at the local, state, or federal levels. However, the model remains valuable for tracking shifts in opioid vulnerability, as external policy changes and other interventions will manifest in its predictions, revealing alterations in vulnerability patterns at multiple geographic scales—including the county, state, and national levels. It provides a critical tool for understanding how the opioid landscape evolves following such changes. Additionally, incorporating spatial correlations allows the model to capture regional patterns that traditional time-series methods like ARIMA or the generic Kalman filter may overlook, underscoring its broader applicability to public health research and spatial epidemiology.}

\rev{Despite its strengths, this study does have certain limitations. The restricted range of years with consistent data across all three datasets posed challenges for capturing longer-term trends. The dispensing rates dataset contained a notable amount of missing data, requiring careful handling. Furthermore, the final year of this study’s time frame coincides with the onset of the COVID-19 pandemic, confounding the pandemic's broader societal effects with the societal drivers of opioid abuse. These effects may limit the comparability of 2020 data to prior years and affect the generalizability of findings from that year.}

\subsection*{Abbreviations}

\noindent CDC: Centers for disease control
\noindent SVI: Social vulnerability index

\subsection*{Acknowledgments}

\noindent The authors would like to thank Andrew Farrell for his helpful comments and feedback, not only on the manuscript, but throughout the entire study. The authors would also like to thank the anonymous reviewers for their helpful comments and suggestions, which helped refine and strengthen the manuscript.

\subsection{Author contributions}

\noindent AD conducted all core research activities, including curating and analyzing the data, generating the code, and writing the initial manuscript. AS conceptualized the study and provided general oversight and expertise. VM proposed the use of the Kalman filter for prediction, supervised the detailed execution of the project and provided methodological expertise. All authors reviewed the manuscript and contributed to revisions.

\subsection*{Funding}

\noindent This work is sponsored by the US Department of Veterans Affairs using resources from the Knowledge Discovery Infrastructure which is located at the Oak Ridge National Laboratory and supported by the Office of Science of the U.S. Department of Energy. This manuscript has been authored by UT-Battelle, LLC, under contract DE-AC05-00OR22725 with the US Department of Energy (DOE). The US government retains and the publisher, by accepting the article for publication, acknowledges that the US government retains a nonexclusive, paid-up, irrevocable, worldwide license to publish or reproduce the published form of this manuscript, or allow others to do so, for US government purposes. DOE will provide public access to these results of federally sponsored research in accordance with the DOE Public Access Plan (\url{http://energy.gov/downloads/doe-public-access-plan})

\subsection*{Availability of data and materials}

\noindent The appendix, datasets and code supporting the conclusions of this article are available in the `A-Deas/Hotspots' Github repository: \url{https://github.com/A-Deas/Hotspots.git}.

\subsection*{Ethics approval and consent to participate}

\noindent Not applicable.

\subsection*{Consent for publication}

\noindent Not applicable.

\subsection*{Competing interests}

\noindent The authors have no competing interests to declare.

\subsection*{Author information}

\noindent \textsuperscript{1}Department of Mathematics, University of Tennessee, Circle Dr, Knoxville, 37916, TN, USA. \textsuperscript{2}The Bredesen Center for Interdisciplinary Research and Graduate Education, University of Tennessee, Middle Dr, Knoxville, 37996, TN, USA. \textsuperscript{3}Computational Sciences and Engineering Division, Oak Ridge National Laboratory, Bethel Valley Road, Oak Ridge, 37830, TN, USA. \textsuperscript{4}Office of Mental Health and Suicide Prevention, Veterans Health Administration,Willow Road, Palo Alto, 94025, CA, USA.

\renewcommand{\refname}{References}
\def\UrlBreaks{\do\/\do-}
\makeatletter\renewcommand\@biblabel[1]{#1.}\makeatother

\section*{Appendix}
\appendix
\renewcommand{\thesection}{\Alph{section}} 

\section{The Kalman filter algorithm}\label{apdx_kalman_algo}

We present a mathematical exposition of the Kalman filter algorithm. The Kalman filter estimates a system's state over time through prediction and update phases. In the prediction phase, the current state, $\hat{N}_{t|t-1}$, and error covariance matrix, $P_{t|t-1}$, are forecast based on the previous state and error covariance matrix as follows:
\begin{align*}
\hat{N}_{t|t-1} &= F_t \hat{N}_{t-1|t-1} \\
P_{t|t-1} &= F_t P_{t-1|t-1} F_t^\intercal + Q_t.
\end{align*}
We set the matrix $Q_t$ to the covariance matrix $Q$ in \cref{eq:state_evolution} for all $t$, which encapsulates the spatial relations in the data defined in \cref{eq:measurement_eqn}. The matrix $F_t$ models the system dynamics, taken to be the identity matrix for all $t$ as our model assumes a linear progression of the system state over time.

During the update phase, upon receiving the annual data $D_t$, the filter refines its estimates using a weighted average term, $K_t$, called the Kalman gain. The Kalman gain is calculated to adjust predictions based these new measurements, subsequently updating the current state estimate $\hat{N}_{t|t}$ and error covariance matrix $P_{t|t}$ as follows:
\begin{align*}
K_t &= P_{t|t-1} H_t^\intercal (H_t P_{t|t-1} H_t^\intercal + R_t)^{-1} \\
\hat{N}_{t|t} &= \hat{N}_{t|t-1} + K_t(D_t - H_t \hat{N}_{t|t-1}) \\
P_{t|t} &= (I - K_t H_t) P_{t|t-1}.
\end{align*}
$H_t$ is the measurement model, taken to be the identity matrix in order to directly identify counties with their data. $R_t$ is the measurement noise covariance matrix, set to the identity matrix scaled by 0.01 to reflect the assumed 1\% uncertainty in the data measurements.

\section{Data uncertainty level sensitivity analysis}\label{apdx_uncertainty}

To choose a data uncertainty level for our study, we conducted a sensitivity analysis. This analysis examined various uncertainty levels, specifically 1\%, 3\%, and 5\%, by evaluating the respective changes in the model's 2020 predictive performance. \rev{Across all datasets, we observed a decline in accuracy as uncertainty increased. Given these findings, we opted to use a 1\% uncertainty level for all data measurements throughout our study to maintain optimal predictive accuracy. \Cref{table_data_uncertainty} summarizes the impact of increasing data uncertainty on the filter’s 2020 average general accuracy.}

\begin{table}[h]
\centering
\begin{tabular}{
    |p{3.5cm}| 
    p{1.5cm}|
    p{1.5cm}|
    p{1.5cm}|
}
 \hline
\multicolumn{1}{|>{\columncolor{lightgrey}}c|}{\textbf{Data Uncertainty Level}} & 
\multicolumn{1}{>{\columncolor{lightgrey}}c|}{\textbf{Mortality}} & 
\multicolumn{1}{>{\columncolor{lightgrey}}c|}{\textbf{Disability}} & 
\multicolumn{1}{>{\columncolor{lightgrey}}c|}{\textbf{Dispensing}}  \\
 \hline\noalign{\vskip 3.5pt}\hline

 \hline
 1\% uncertainty & 94.00\% & 88.49\% & 91.61\% \\
 3\% uncertainty & 93.87\% & 87.97\% & 89.96\% \\
 5\% uncertainty & 93.81\% & 87.70\% & 88.99\% \\
 
\hline\noalign{\vskip 3.5pt}
\end{tabular}
\caption{The affects on the filter's 2020 general accuracy as the level of uncertainty in the data increases.}
\label{table_data_uncertainty}
\end{table}

\section{Multi-year predictive efficacy of the spatial Kalman filter}\label{apdx_told_comparison}

Our study optimized the Kalman filter's predictive accuracy by using all five available training years from 2015 to 2019, then forecasting only for 2020. However, the filter is capable of making predictions over multiple years. Here we explore the multi-year predictive efficacy of the filter by training it with progressively less data, decreasing from four training years down to just one, then extending the prediction periods accordingly. For example, the filter trained on a single year of data, is initialized with 2014 data, learns from 2015 data then generates predictions for the years 2016 through 2020. This approach allowed us to assess the impact of training length on prediction accuracy over extended forecasting horizons.

\begin{figure*}[p]
    \centering
    \includegraphics[width=\linewidth]{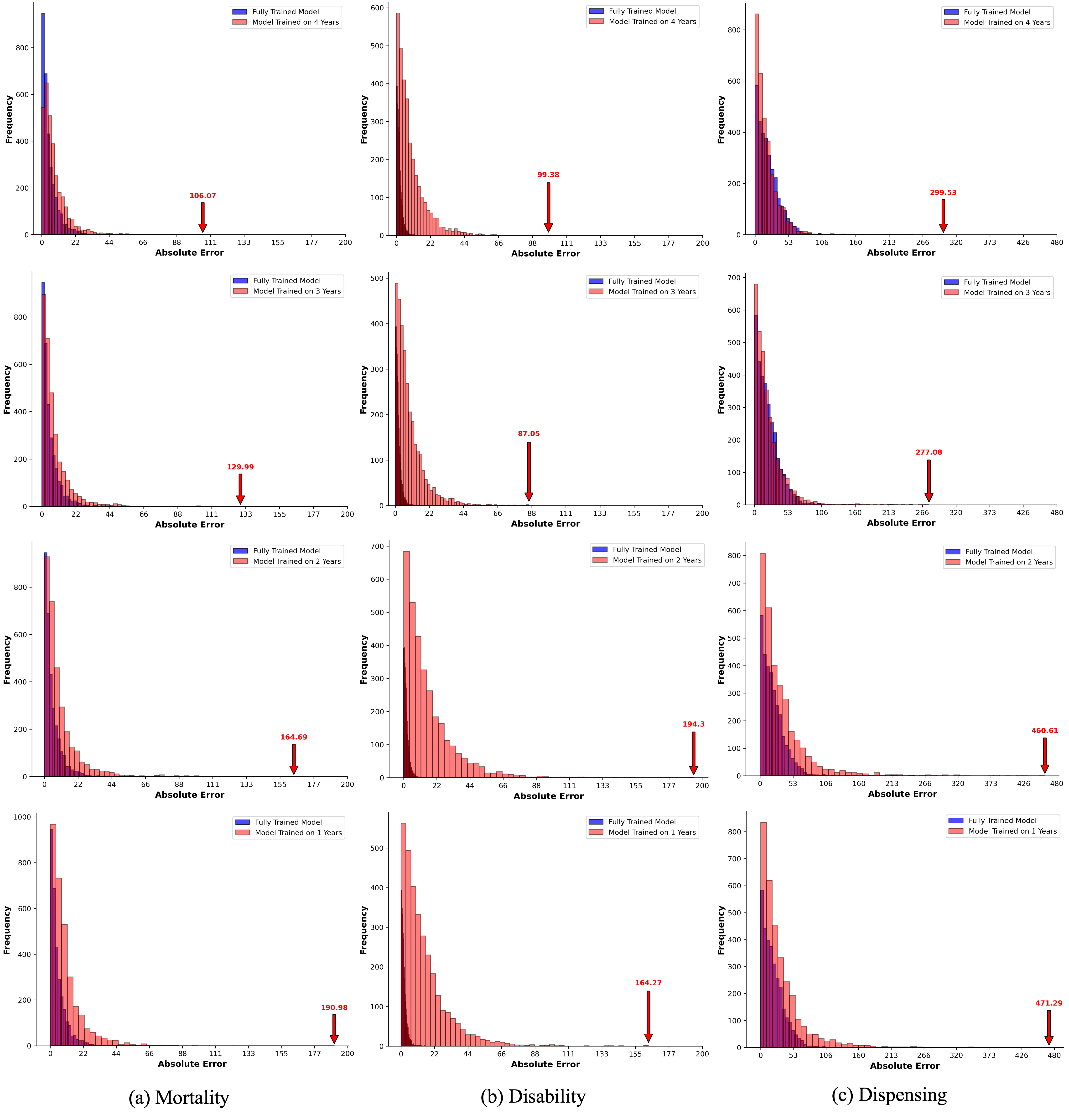}
    \caption{Histogram comparisons of the absolute errors for the 2020 (a) mortality, (b) disability, and (c) dispensing rate predictions between the fully trained filter and the filter trained on progressively less data. The error distribution of the fully trained filter is colored in blue, whereas the error distribution of the filter trained on less data is colored in red. The top row showcases comparisons for the filter trained on four years of data, iteratively progressing downwards to the last row showcasing comparisons for the filter trained on a single year of data. Each histogram features a red arrow highlighting the maximum error from the filter trained on less data. Figures are best viewed online in color.}
    \label{multi_year_err_comparison}
\end{figure*}

\Cref{multi_year_err_comparison} showcases a comparative histogram analysis of the filter's errors for the year 2020 when it is trained on progressively less data versus the fully trained filter. Since errors steadily accumulate for each consecutive prediction year, we showcase the results for 2020 to offer a clear view of the filter’s performance degradation as it is trained on less data. The histogram comparisons for the remaining years in the study can be found in the supplemental materials. 

As expected, training the filter on less data leads to an incremental increase in both the frequency of errors and the magnitude of the maximum error, the latter of which is highlighted by a red arrow in each histogram. As the training period shortens for mortality and dispensing rates, we note an increase in maximum errors and the error distributions shifting rightward. Despite this, considering the limited data it had to learn from, the filter still maintains a commendable performance on these datasets.

However, the biennial publication of disability rates, coupled with our method of synthesizing the intervening annual data, poses distinct challenges for multi-year disability rate predictions. Specifically, this data structure causes the filter to mistakenly extrapolate trends over two-year periods, leading to considerable errors in its multi-year predictions. Such issues become intractable in the absence of intervening data points and result in the filter failing to properly adjust its projections. This problem is evident in \cref{multi_year_err_comparison}, where the error distributions of the filter trained on less data are notably more extensive and pronounced compared to the fully trained filter. This scenario underscores the critical importance of consistent annual data collection to sustain the filter's predictive accuracy across longer forecasting intervals.

\newpage
\section{Impact of Initialization Year on Prediction Errors}\label{apdx_initializations_years}

\rev{In the previous section, we examined how reducing the Kalman filter's training period while extending the forecasting horizon affected predictive errors. Here, we instead evaluate how shifting the initialization year affects the model's 2020 predictions. Given the dynamic nature of opioid-related issues, the choice of historical data used to initialize the model may play a crucial role in capturing shifting trends. This analysis provides insight into the model’s adaptability to sudden changes in the data. If a significant shift occurs in opioid-related mortality, prescribing patterns, or social vulnerability factors, an extended training period may cause the model to overemphasize outdated trends rather than adjust to more recent dynamics. By iteratively shifting the initialization year forward, we assess whether relying on more recent observations enhances prediction accuracy or if a longer historical perspective remains beneficial.}

\begin{figure*}[p]
    \centering
    \includegraphics[width=\linewidth]{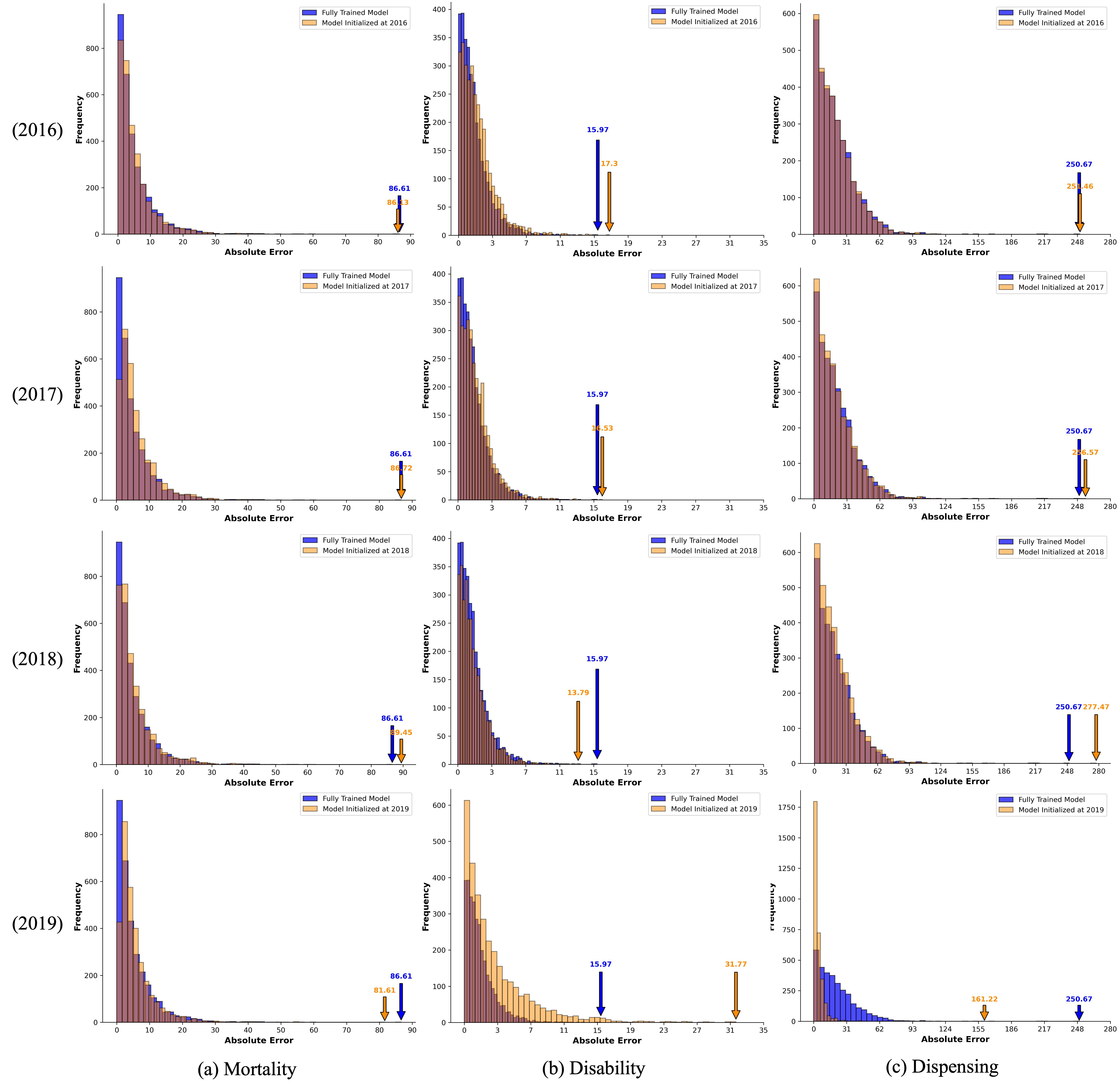}
    \caption{\rev{Histogram comparisons of the absolute errors for the 2020 (a) mortality, (b) disability, and (c) dispensing rate predictions between the fully trained filter and the filter initialized at later years. The error distribution of the fully trained filter is colored in blue, whereas the error distribution of the filter initialized at later years is colored in orange. The top row showcases comparisons with the filter initialized at 2016, the second row for the filter initialized at 2017, the third row for the filter initialized at 2018, and the last row for the filter initialized at 2019. Each histogram features a blue arrow highlighting the maximum error from fully trained Kalman filter and an orange arrow highlighting the maximum error from the filter initialized at later years. Figures are best viewed online in color.}}
    \label{init_year_err_comparison}
\end{figure*}

\rev{\Cref{init_year_err_comparison} presents the absolute error distributions for 2020 as a function of the initialization year, ranging from 2016 to 2019. In nearly all cases, the fully trained model—initialized with data from 2014 and trained through 2019—yields the more accurate results. However, there is one notable exception: When predicting 2020 opioid dispensing rates, the model initialized with 2019 data outperforms the fully trained model by a significant margin. As discussed earlier, major policy changes in 2019 led to a substantial drop in opioid prescriptions from 2018 to 2019. By contrast, from 2019 to 2020, dispensing rates remained stable without any dramatic shifts. The Kalman gain used in the fully trained model incorporates information about the sharp decline from 2018 to 2019, whereas the model initialized in 2019 does not, allowing it to better capture the stability of prescription rates from 2019 to 2020.}

\rev{This result highlights an important consideration when selecting historical training data for time-series forecasting models: Abrupt shifts in a dataset can make older data less relevant—or even misleading—for future predictions. In stable data environments, a longer training period generally provides the best predictive performance by leveraging more comprehensive historical patterns. However, if a dataset contains abrupt changes from one time point to the next, due to policy shifts or other external factors, shorter training horizons may be preferable to avoid distortions caused by outdated trends.}

\rev{Ultimately, our results suggest that shifting the initialization year can be a valuable strategy in contexts where sudden structural changes occur. Practitioners using Kalman filters or similar forecasting models should carefully weigh the trade-off between historical data depth and recent trend sensitivity. If the goal is to capture long-term trends, an extended training period is likely optimal. However, when significant, known disruptions exist, retraining the model on a more recent subset of data may improve short-term forecast accuracy.} 

\newpage
\section{Efficacy summary for the spatial Kalman filter during training years}\label{apdx_training_efficacy}

\begin{table}[h]
\centering
\begin{tabular}{
    |p{3.5cm}| 
    p{1.5cm}|
    p{1.5cm}|
    p{1.5cm}|
    p{1.5cm}|
    p{1.5cm}| 
}
 \hline
\multicolumn{1}{|>{\columncolor{lightgrey}}c|}{\textbf{Variable}} & 
\multicolumn{1}{>{\columncolor{lightgrey}}c|}{\textbf{Year}} & 
\multicolumn{1}{>{\columncolor{lightgrey}}c|}{\textbf{Avg General Acc}} & 
\multicolumn{1}{>{\columncolor{lightgrey}}c|}{\textbf{Hotspot Acc}} & 
\multicolumn{1}{>{\columncolor{lightgrey}}c|}{\textbf{Avg Error}} & 
\multicolumn{1}{>{\columncolor{lightgrey}}c|}{\textbf{Max Error}} \\
 \hline\noalign{\vskip 3.5pt}\hline

 \hline
 Mortality rates & 2015 & 96.57\% & 97.31\% & 0.31 & 9.13 \\
                 & 2016 & 94.52\% & 97.04\% & 0.37 & 6.81 \\
                 & 2017 & 94.25\% & 95.54\% & 0.41 & 7.07 \\
                 & 2018 & 94.49\% & 94.34\% & 0.38 & 6.85 \\
                 & 2019 & 93.83\% & 95.0\% & 0.38 & 6.15 \\
                 & 2020 & 94.00\% & 70.00\% & 5.20 & 86.61 \\
 
\hline\noalign{\vskip 3.5pt}\hline

 Disability rates & 2015 & 92.25\% & 91.49\% & 0.58 & 7.49 \\
                  & 2016 & 91.71\% & 89.66\% & 0.77 & 9.30 \\
                  & 2017 & 92.04\% & 95.45\% & 0.72 & 9.07 \\
                  & 2018 & 92.71\% & 91.86\% & 0.86 & 11.78 \\
                  & 2019 & 91.31\% & 93.81\% & 0.81 & 9.30 \\
                  & 2020 & 88.49\% & 94.19\% & 1.84 & 15.97 \\
 \hline

\hline\noalign{\vskip 3.5pt}\hline

 Dispensing rates& 2015 & 96.47\% & 99.38\% & 0.46 & 13.16 \\
                 & 2016 & 96.05\% & 98.82\% & 0.55 & 13.97 \\
                 & 2017 & 96.53\% & 99.41\% & 0.46 & 13.26 \\
                 & 2018 & 97.99\% & 99.41\% & 0.46 & 23.00 \\
                 & 2019 & 97.50\% & 97.59\% & 1.81 & 72.32 \\
                 & 2020 & 91.61\% & 68.64\% & 21.02 & 250.67 \\
 \hline
\end{tabular}
\caption{Efficacy metric results for the spatial Kalman filter for all predictive years in study}
\label{table_training_summary}
\end{table}

In \cref{table_training_summary}, we provide the efficacy metric results for each dataset and training year in the study. Excluding the initialization year of 2014, the subsequent training years, 2015 to 2019, are pivotal for the filter's learning phase, where its estimates are refined using the observed data. The output for these years therefore consists of calibrated estimates which showcase the filter's evolution as it iteratively learns underlying trends in the data.

\newpage
\section{Data Summary Statistics}\label{apdx_data_summary}

\begin{table}[h]
\centering
\begin{tabular}{
    |p{3.5cm}| 
    p{1cm}|
    p{1cm}|
    p{1cm}|
    p{1cm}|
    p{1cm}|
    p{1cm}|
    p{1cm}|
    p{1cm}| 
}
 \hline
\multicolumn{1}{|>{\columncolor{lightgrey}}c|}{\textbf{Variable}} & 
\multicolumn{1}{>{\columncolor{lightgrey}}c|}{\textbf{Year}} & 
\multicolumn{1}{>{\columncolor{lightgrey}}c|}{\textbf{Mean}} & 
\multicolumn{1}{>{\columncolor{lightgrey}}c|}{\textbf{Std Dev}} & 
\multicolumn{1}{>{\columncolor{lightgrey}}c|}{\textbf{Min}} & 
\multicolumn{1}{>{\columncolor{lightgrey}}c|}{\textbf{Q1}} & 
\multicolumn{1}{>{\columncolor{lightgrey}}c|}{\textbf{Median}} & 
\multicolumn{1}{>{\columncolor{lightgrey}}c|}{\textbf{Q3}} & 
\multicolumn{1}{>{\columncolor{lightgrey}}c|}{\textbf{Max}} \\
 \hline\noalign{\vskip 3.5pt}\hline

 \hline            
 Mortality rates & 2014 & 13.55 & 8.01 & 0.00 & 8.10 & 11.80 & 16.90 & 87.90 \\
                 & 2015 & 14.84 & 8.83 & 0.00 & 8.90 & 12.90 & 18.50 & 113.90 \\
                 & 2016 & 17.07 & 10.29 & 0.00 & 10.00 & 14.40 & 21.30 & 94.80 \\
                 & 2017 & 18.46 & 11.42 & 0.00 & 10.80 & 15.60 & 23.00 & 141.50 \\
                 & 2018 & 17.53 & 10.50 & 0.00 & 10.30 & 14.90 & 21.85 & 114.40 \\
                 & 2019 & 18.01 & 10.78 & 0.00 & 10.70 & 15.40 & 22.30 & 117.30 \\
                 & 2020 & 23.54 & 14.61 & 0.00 & 13.70 & 20.00 & 29.20 & 153.20 \\
 
\hline\noalign{\vskip 3.5pt}

 \hline            
Disability rates & 2014 & 49.64 & 28.94 & 0.00 & 24.45 & 48.93 & 74.59 & 99.97 \\
                 & 2015 & 49.64 & 28.60 & 0.00 & 24.97 & 49.48 & 74.20 & 99.94 \\
                 & 2016 & 49.63 & 28.95 & 0.00 & 24.62 & 49.38 & 74.50 & 100.00 \\
                 & 2017 & 49.63 & 28.56 & 0.00 & 25.39 & 50.02 & 74.19 & 99.94 \\
                 & 2018 & 49.63 & 28.96 & 0.00 & 24.83 & 49.06 & 74.34 & 100.00 \\
                 & 2019 & 49.62 & 28.43 & 0.00 & 25.46 & 49.44 & 74.26 & 99.97 \\
                 & 2020 & 49.62 & 28.96 & 0.00 & 24.47 & 49.49 & 74.63 & 100.00 \\
 \hline

\hline\noalign{\vskip 3.5pt}

 \hline            
Dispensing rates & 2014 & 80.59 & 51.90 & 0.00 & 47.70 & 79.40 & 110.55 & 563.30 \\
                 & 2015 & 75.10 & 47.88 & 0.00 & 45.60 & 73.50 & 103.30 & 504.90 \\
                 & 2016 & 71.91 & 44.57 & 0.00 & 44.25 & 70.50 & 97.65 & 470.30 \\
                 & 2017 & 64.61 & 40.13 & 0.00 & 39.50 & 62.70 & 87.55 & 402.00 \\
                 & 2018 & 56.65 & 35.70 & 0.00 & 34.60 & 54.90 & 76.25 & 311.30 \\
                 & 2019 & 40.49 & 33.11 & 0.00 & 19.55 & 34.60 & 54.00 & 567.90 \\
                 & 2020 & 38.01 & 31.08 & 0.00 & 17.80 & 32.30 & 51.00 & 406.70 \\
 \hline
\end{tabular}
\caption{\rev{Data summary statistics from 2014 to 2020.}}
\label{table_data_summary_statistics}
\end{table}

\section{Comparison of the Spatial and Generic Kalman Filters}\label{apdx_generic_comparison}

\begin{figure*}[h]
    \centering
    \includegraphics[width=\linewidth]{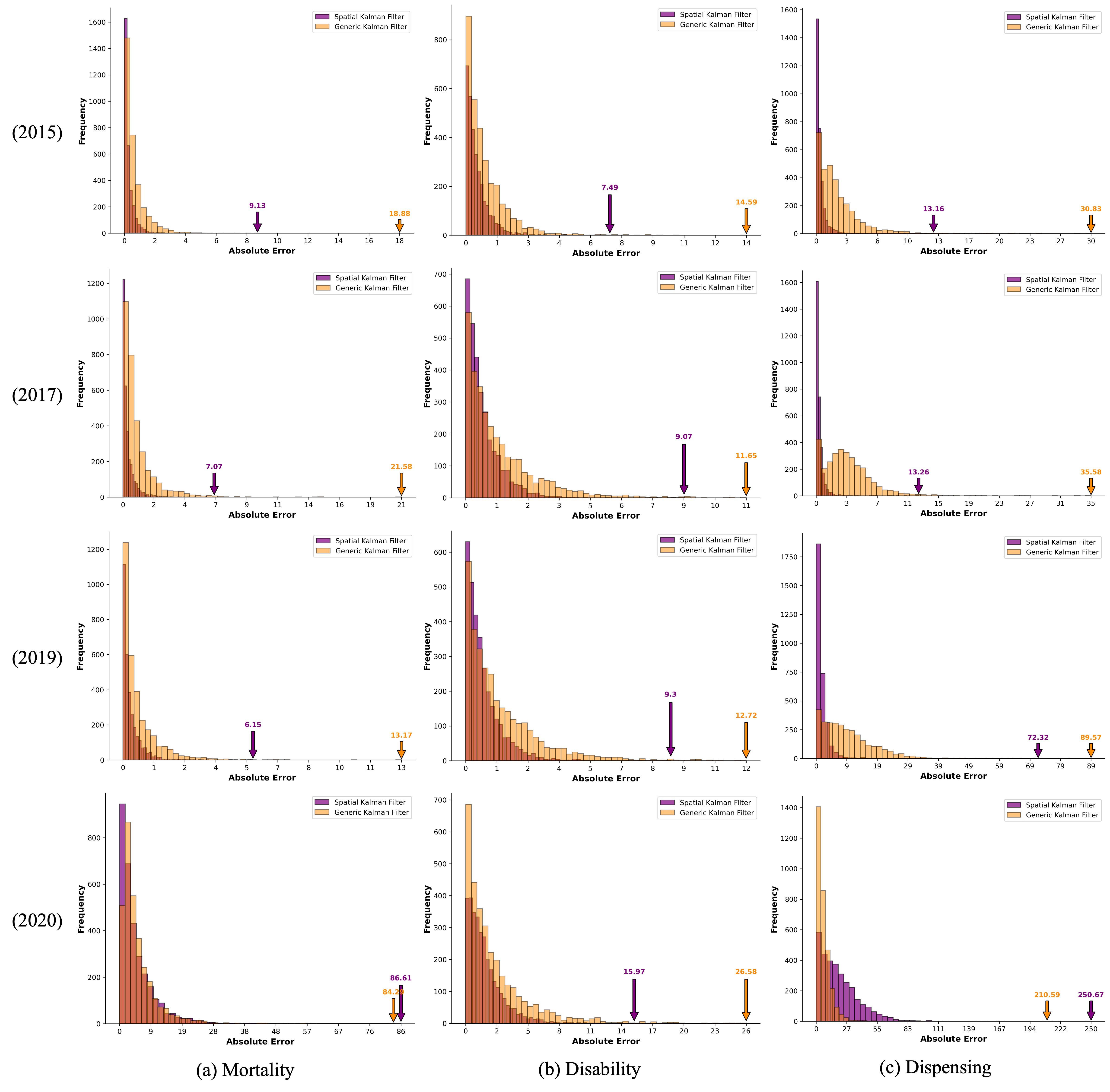}
    \caption{\rev{Histogram comparisons of the spatial and generic kalman filter's absolute errors for the 2015, 2017, 2019 and 2020 (a) mortality, (b) disability, and (c) dispensing rate predictions. The error distributions of the spatial Kalman filter are colored in purple, whereas the generic Kalman filter's error distributions are colored in orange. The top row showcases comparisons for year 2015, the second row for 2017, the third row for 2019, and the last row for 2020. Each histogram features a purple arrow highlighting the maximum error from spatial Kalman filter and an orange arrow highlighting the maximum error from the generic Kalman filter. Figures are best viewed online in color.}}
    \label{fig_generic_comparison}
\end{figure*}

\end{document}